\documentstyle[twocolumn,graphics,prl,aps,floats]{revtex}

\newcommand{\lu}{LuNi$_2$B$_2$C}

\begin{document}

\wideabs{
\draft
\title{
Highly Anisotropic Gap Function in Borocarbide Superconductor LuNi$_2$B$_2$C}

\author{Etienne Boaknin, R.W. Hill,
Cyril Proust, C. Lupien and Louis Taillefer}

\address{Canadian Institute for Advanced Research}

\address{
Department of Physics, University of Toronto, Toronto,
Ontario M5S 1A7, Canada}

\author{P.C. Canfield}

\address{Ames Laboratory, Department of Physics and Astronomy, Iowa
  State University, Ames, Iowa 50011}

\date{\today}

\maketitle

\begin{abstract}
  The thermal conductivity of borocarbide superconductor
  LuNi$_2$B$_2$C was measured down to 70 mK ($T_c/200$) in a magnetic
  field perpendicular to the heat current from $H =0$ to above $H_{c2}
  = 7$~T.  As soon as vortices enter the sample, the conduction at $T
  \to 0$ grows rapidly, showing unambiguously that delocalized
  quasiparticles are present at the lowest energies.  The field
  dependence is very similar to that of UPt$_3$, a heavy-fermion
  superconductor with a line of nodes in the gap, and very different
  from the exponential dependence characteristic of $s$-wave
  superconductors.  This is strong evidence for a highly anisotropic
  gap function in LuNi$_2$B$_2$C, possibly with nodes.

\end{abstract}

}

The vast majority of known superconductors are characterized by an
order parameter with $s$-wave symmetry and a gap function which is
largely isotropic and without nodes (zeros).  Only four families of
materials are seriously thought to exhibit a superconducting state
with a different symmetry: (1) heavy-fermion materials, such as UPt$_3$
where a line of nodes in the gap function has clearly been identified
\cite{UPt3}; (2) high-$T_c$ cuprates, such as YBa$_2$Cu$_3$O$_7$ where
the order parameter was clearly shown to have $d$-wave symmetry
\cite{Cuprates}; (3) the ruthenate Sr$_2$RuO$_4$, where there is strong
evidence for a triplet order parameter \cite{Ruthenates}; and (4)
organic conductors, such as $\kappa$-(ET)$_2$Cu[N(CN)$_2$]Cl where
there is growing evidence for unconventional superconductivity
\cite{Organics}.  A major outstanding question is the nature of the
microscopic mechanism responsible for superconductivity in any of
these materials.  The unconventional symmetry of the order parameter
is evidence for a pairing caused by purely electronic interactions and
not mediated by phonons.  For example, the proximity to magnetic order
which is found in all four families of superconductors has led to the
suggestion that spin fluctuations are responsible for Cooper pairing,
as is thought to be the case in superfluid $^3$He.

The presence of nodes in the gap function is generally associated with
unconventional (non-$s$-wave) symmetries. These nodes are typically
inferred from the observation of quasiparticle excitations at energies
much lower than the gap maximum $\Delta_0$, as reflected for example
in the power law temperature dependence of various physical
properties, such as London penetration depth and ultrasonic
attenuation at $T \ll T_c$.  Another way of detecting low-energy
quasiparticles is to excite them by applying a magnetic field which
introduces vortices in the material, so that the superfluid flow
around each vortex Doppler shifts the quasiparticle energy.  In
certain limits, the quasiparticle response is the same whether induced
by a thermal energy $k_B T$ or by a field energy $\simeq \Delta_0
\sqrt{B/B_{c2}}$, where $B_{c2} \simeq H_{c2}$, the upper critical field
\cite{Simon}.

In this Letter, we turn our attention to another class of
superconductors: the borocarbides {\sl L}Ni$_2$B$_2$C (where {\sl L} =
Y, Lu, Tm, Er, Ho, and Dy) \cite{Physics-Today}.  It has generally been
thought that these materials are described by an order parameter with
$s$-wave symmetry and pairing which proceeds via the electron-phonon
coupling \cite{Carter,Mattheiss,Eliashberg}.  However, there is recent
evidence for low-energy excitations in the superconducting state,
whether from the anomalous field dependence of the specific heat
\cite{Nohara,Izawa} and the microwave surface impedance
\cite{Izawa,Jacobs},
or from the presence of scattering below the gap in Raman measurements
\cite{raman}.  This has been interpreted in terms of an anisotropic
$s$-wave gap \cite{Nohara,Izawa} (see also reference \cite{Yokoya}).

Here we present compelling evidence that the gap function of
LuNi$_2$B$_2$C is highly anisotropic, with a gap minimum $\Delta_{\rm
  min}$ at least 10 times smaller than the gap maximum, $\Delta_{\rm
  min} \leq \Delta_0/10$, and possibly going to zero at nodes.  This
statement is based on the observation of delocalized quasiparticles at
very low energies, as measured directly by heat transport.  Indeed,
quasiparticle conduction is induced by a magnetic field as low as
$H_{c1} \simeq H_{c2}/100$ and it grows linearly with field, in
dramatic contrast with the exponentially activated transport seen in
Nb, for example, where it results from tunneling between the localized
states bound to the core of adjacent vortices.  Such pronounced
anisotropy challenges the current view on the nature of
superconductivity in borocarbides. It suggests either a new family of
unconventional superconductors (with symmetry-imposed nodes in the gap
function) or $s$-wave superconductors with more anisotropy than has
ever been seen before. In either case, the role of phonons as a
proposed pairing mechanism may have to be critically re-examined.

LuNi$_2$B$_2$C is an extreme type-II superconductor and a non-magnetic
member of the borocarbide family with a superconducting transition
temperature $T_c \simeq 16$~K and an upper critical field
$H_{c2}(0)\simeq 7$~T.  Its thermal conductivity $\kappa$ was measured
in a dilution refrigerator using a standard steady-state technique.  A
heater and two RuO$_2$ thermometers were used for the measurements.
The later were calibrated in-situ for each field against a Germanium
thermometer. The temperature was increased at fixed magnetic field,
from 70 mK up, in fields ranging from 0 to 8 T.  The field was applied
parallel to the $c$-axis of the tetragonal crystal structure ([001]),
and perpendicular to the heat current (along [100]).  Below 1.5 T, the
sample was cooled in the field to ensure a good field homogeneity.
Above 1.5 T, the results were independent of cooling procedure.  The
single crystal was grown by a melting flux method
\cite{Physics-Today}.  The sample was a rectangular parallepiped of
width 0.495 mm (along [001]) and thickness 0.233 mm (along [010]),
with a 1.59 mm separation between contacts (along [100]).
\begin{figure}
\resizebox{\linewidth}{!}{\includegraphics{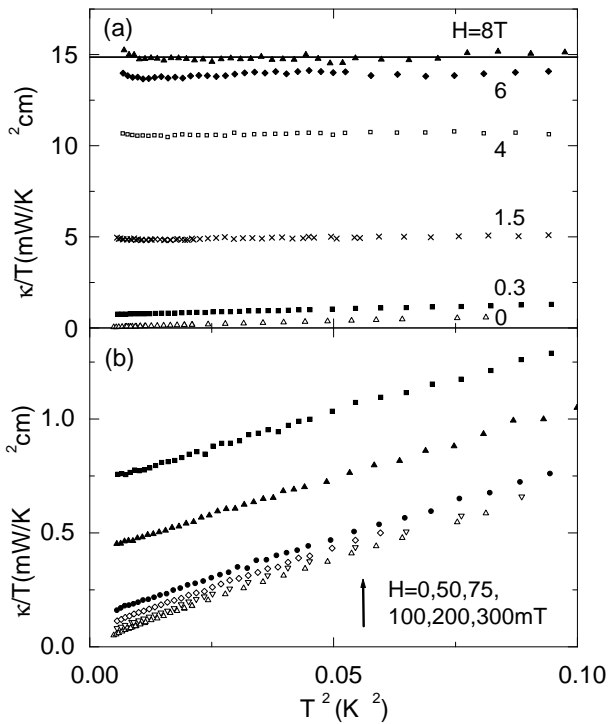}}
\caption{Temperature dependence of thermal conductivity at several
  applied fields,
  plotted as $\kappa/T$ vs $T^2$, for
  (a) $H$~= 0,
  0.3, 1.5, 4, 6 and 8 T, and
  (b) $H$~ = 0, 50, 75, 100, 200 and 300~mT, in
  increasing order.  
  The solid line indicates the
  value expected from the Wiedemann-Franz law above $H_{c2}$. }
\end{figure}
\noindent
In zero magnetic field, $\rho$(300 K) = 35 $\mu \Omega$ cm and
$\rho_0$ = 1.30~$\mu \Omega$ cm, from a fit to $\rho$ = $\rho_0 +
AT^2$ between $T_c$ and 50 K.
There is a positive magnetoresistance such that $\rho$(8~T) = 1.67
$\mu \Omega$ cm at $T \to 0$.  The zero-temperature coherence length
$\xi_0 = 70$~\AA, using $H_{c2}(0) = \Phi_0 / 2 \pi \xi_0^2$.  The
zero temperature penetration depth is $\lambda_0 = 760$~\AA
\cite{Rathnayaka}.  The mean free path is approximately 500 \AA.  We
measured the lower critical field $H_{c1}(0)$ to be 60~mT, using the
sudden drop in $\kappa(H)$ vs $H$ at 2~K, caused by the strong
scattering of phonons by vortices as they first enter the sample.

The thermal conductivity $\kappa(T)$ of \lu\ is plotted in Fig.~1, as
$\kappa/T$ vs $T^2$. The total conductivity is the sum of an
electronic and a phononic contribution: $\kappa = \kappa_e +
\kappa_{ph}$.  By plotting the data in this way, one can easily
separate the electronic term linear in $T$ from the phononic term
cubic in $T$.  As the temperature is decreased, and in the absence of
strong electron-phonon scattering ({\it i.e.} at $H \simeq 0$), the
phonon mean free path eventually grows to reach the size of the
crystal, at which point $\kappa_{ph} \sim T^3$.  The low-field curves
(in the lower panel of Fig.~1) are indeed roughly linear (and
parallel) in such a plot, with a slope in quantitative agreement with
the known sound velocities and sample dimensions, as reported earlier
\cite{Boaknin}.  At higher field, roughly above 2~T and all the way
into the normal state, the conduction is essentially entirely due to
electrons and
given by a constant $\kappa/T$. Note that the magnitude of this linear
$\kappa$ is in perfect agreement with the Wiedemann-Franz law, namely
$\kappa_e/T =L_0/\rho$, where $L_0 = (\pi^2/3) (k_B/e)^2=2.45 \times
10^{-8}$~W~$\Omega$~K$^{-2}$ and $\rho$ = $\rho$(8~T).  Given this
well-understood behaviour of $\kappa_{e}(T)$ and $\kappa_{ph}(T)$, it
is straightforward to extract the electronic contribution
$\kappa_e(T)$, by simply extrapolating $\kappa/T$ to $T = 0$.  The
result of this extrapolation is plotted as $\kappa_e/T$ vs $H$ in
Figs.~2 and 3, where the field is normalized to unity at $H_{c2}(0)$
and the conductivity, to its normal state value.  One immediately
notices the large amount of {\it delocalized} quasiparticles
throughout the vortex state of \lu.  [This would seem to provide a
natural explanation for the observation of de Haas-van Alphen
oscillations down to unusually low fields ($H_{c2}/5$) in
YNi$_2$B$_2$C \cite{dHvA-boro}, a close cousin of \lu, with
$T_c=15.5$~K and $H_{c2}=6.5$~T.]  In fact, the growth of quasiparticle
conduction starts right at $H_{c1}$ and is seen to be roughly linear
in field. This is in dramatic contrast with the behaviour of
quasiparticles in $s$-wave superconductors with a large finite gap for
all directions of electron motion.  For comparison, we show in Fig.~2
the electronic conductivity of Nb measured at 2~K ({\it i.e.}
0.22~$T_c$) \cite{Lowell}.  In an isotropic $s$-wave superconductor,
the only quasiparticle states present at $T \ll T_c$ are those
associated with vortices. When vortices are far apart, these states
are bound to the vortex core and are therefore localized, and unable
to transport heat.  They thus contribute to the specific heat but not
to the thermal conductivity.  As the field is increased and the
vortices are brought closer together, tunneling between states on
adjacent vortices will cause some delocalization.  This conduction is
expected to grow exponentially with the ratio of intervortex
separation to vortex core size ($\simeq \xi_0$), namely as exp($-
\alpha \sqrt{H_{c2}/H}$), where $\alpha$ is a constant, as is found
for Nb at fields below $H_{c2}/3$ \cite{Vinen}.

In the presence of nodes in the gap, the dominant mechanism for
quasiparticle transport in the vortex state is totally different.
Conduction results from the population of extended quasiparticle
states in the bulk of the sample, {\it outside} the vortex cores. The
excitation of these quasiparticles proceeds via the Doppler shift of
their energies as they move in the presence of the superfluid flow
circulating around each vortex. Because near the nodes such states
exist down to zero energy, the growth in the zero-energy quasiparticle
density of states starts right at $H_{c1}$, with a characteristic
$\sqrt{H}$ dependence \cite{Volovik}. This leads to a $\sqrt{H}$
dependence of the specific heat at low temperature, as observed for
example in the cuprate superconductor YBa$_2$Cu$_3$O$_7$ \cite{Moler}.
\begin{figure}
  \resizebox{\linewidth}{!}{\includegraphics{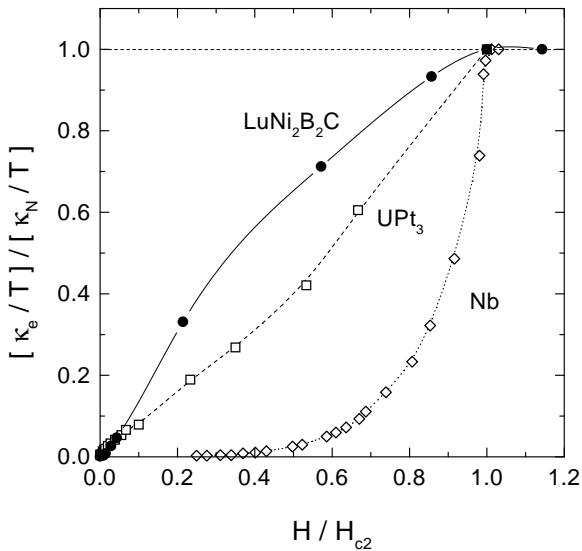}}
\caption{Magnetic field dependence of the electronic thermal conductivity
  $\kappa/T$ at $T \to 0$, normalized to its value at $H_{c2}$.
  Circles are for LuNi$_2$B$_2$C, squares for UPt$_3$
  \protect\cite{Suderow} and diamonds for Nb \protect\cite{Lowell}.
  Note the qualitative difference between the activated conductivity
  of $s$-wave superconductor Nb and the roughly linear growth seen in
  UPt$_3$, a superconductor with a line of nodes. The lines are a
  guide to the eye.}
\end{figure}
Note that the same mechanism will operate for an anisotropic $s$-wave
gap if the field is such that the Doppler shift exceeds the minimum
gap in the quasiparticle spectrum.  It is worth noting, however, that
a similar field dependence has also been observed in $s$-wave
superconductors, such as NbSe$_2$ \cite{Sonier} where it has been
attributed to the bound states in the vortex core.  Specific heat
studies are therefore unable to distinguish between a $\sqrt{H}$
contribution coming from localized core states and that coming from
extended states outside the core.  In contrast, thermal conductivity
is selective, in that it only probes the contribution of delocalized
excitations.

The effect of vortices on quasiparticle transport in an unconventional
superconductor with a line of nodes in the gap function was studied in
beautiful detail by Suderow and co-workers \cite{Suderow}. Their
measurements of $\kappa(T,H)$ in
UPt$_3$ yield a roughly linear increase of $\kappa_e/T$ at $T \to 0$
\begin{figure}
  \resizebox{\linewidth}{!}{\includegraphics{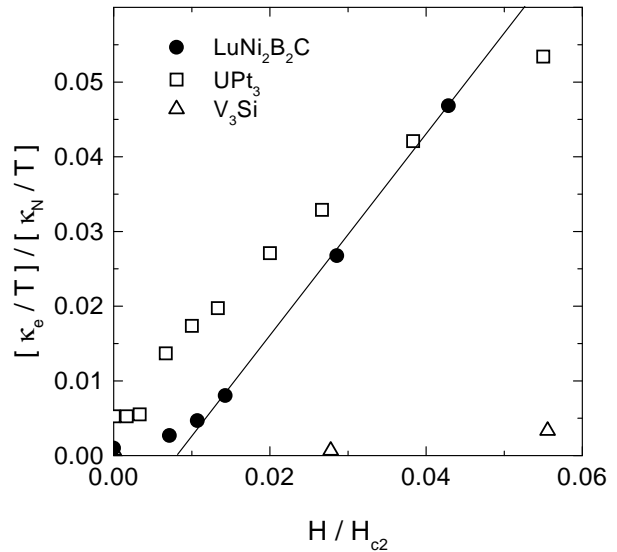}}
\caption{Field dependence of the electronic thermal conductivity
  $\kappa/T$ at $T \to 0$ at low fields, normalized to its value at
  $H_{c2}$. For LuNi$_2$B$_2$C, the growth is linear and starts at
  $H_{c1}$, as emphasized by the solid line. The growth is equally
  rapid for UPt$_3$ \protect\cite{Suderow}.  The equivalent data for $s$-wave
  superconductor V$_3$Si shows a much slower growth.}
\end{figure}
with $H$, shown in Figs.~2 and 3.  The data is for a heat current in
the basal plane of the hexagonal crystal structure, which probes the
equatorial line node in the gap function of UPt$_3$, established by
transverse ultrasound attenuation \cite{Shivaram}.  {\it Figs. 2 and 3
  reveal that quasiparticle conduction in the basal plane of \lu\ is
  as good as in UPt$_3$ (or even better)}.  At low fields, the growth
in the residual linear term $\kappa_0/T \equiv (\kappa/T)_{T \to 0}$
is also linear in $H$, starting at $H_{c1}$:

\begin{equation}
\frac{\kappa_0}{T} \simeq \frac{L_0}{\rho_0} \frac{H - H_{c1}}{H_{c2}}~~~,
\end{equation}

where $\rho_0$ is the zero-field normal-state resistivity.  This is
vastly more conductive than a typical $s$-wave superconductor.  For
example, electronic conduction in V$_3$Si, an extreme type-II $s$-wave
superconductor with comparable $T_c$ (16.5~K) and $\xi_0$ (45 \AA), is
20 times weaker at $H = 0.05~H_{c2}$, as seen from data shown in
Fig.~3.

In both LuNi$_2$B$_2$C and UPt$_3$, the thermal conductivity is
roughly linear in $H$ and the heat capacity follows approximately a
$\sqrt{H}$ dependence. The latter is naturally understood in terms of
a density of states which is linear in energy (coming from nodes or
minima).  However, a theory that can successfully account for the
linear field dependence of $\kappa_e/T$ has not yet been formulated.

Nohara {\it et al.} \cite{Nohara} and Izawa {\it et al.} \cite{Izawa}
have recently attributed the $\sqrt{H}$ dependence of the specific
heat they observe in YNi$_2$B$_2$C to a Doppler shift of the
quasiparticle spectrum as in a $d$-wave superconductor \cite{Kubert}
but applied in this case to a highly anisotropic $s$-wave gap, with a
small minimum gap $\Delta_{\rm min}$.  Interpreting the thermal
conductivity data in the same way yields an estimate of $\Delta_{\rm
  min}$.  Indeed, because quasiparticle conduction starts right at
$H_{c1}$, the minimum gap must be smaller than the Doppler shift
energy $E_H$ at $H_{c1}$.  In a superconductor with a line of nodes in
the gap, the average $E_H$ is given by $ \simeq \Delta_0
\sqrt{B/B_{c2}}$ \cite{Kubert} where $B$ is the magnetic field inside
the superconductor and $B_{c2} \simeq H_{c2}$. At $H_{c1}$, $B\simeq0$
in a type II superconductor thus possibly implying a true zero in the
gap. A conservative upper bound on the minimum field required to
excite quasiparticles above $\Delta_{\rm min}$ uses $B=H_{c1}$. This
gives:

\begin{equation}
\Delta_{\rm min} \leq E_H(H_{c1}) \simeq \Delta_0 \sqrt{H_{c1}/H_{c2}} \simeq \Delta_0 / 10 ~~.
\end{equation}
In other words, there is a huge gap anisotropy,
with a minimum in the basal plane (the direction
of heat current).

A factor 10 in gap anisotropy is unprecedented for an $s$-wave
superconductor, with a factor of 2 being the most that has ever been
inferred in elemental superconductors \cite{Anisotropy}.  Faced with
this striking result, two questions arise: (1) does the gap function
have $s$-wave symmetry (with deep minima) or rather another symmetry
(with actual nodes)?  (2) is the pairing indeed due to phonons?

In relation to the first question, we stress that no sizable residual
linear term $\kappa_0/T$ is observed in $H=0$ (see \cite{Boaknin}), a
fact which would tend to argue against the presence of nodes in the
superconducting gap.  However, this may be a question of magnitude. In the
cuprates, a value of $\kappa_0/T$ in excellent quantitative agreement
with theory has been observed \cite{Chiao}. On the other hand, in
UPt$_3$ \cite{Suderow} the observed $\kappa_0/T$ is significantly
smaller than expected (and in fact barely resolvable)
even though there is overwhelming evidence for
nodes.

A possible test of the symmetry of the order parameter and the nature
of potential nodes (imposed by symmetry vs accidental) is to
investigate the effect of adding impurities. While impurity scattering
will reduce the anisotropy of an $s$-wave gap (either by removing
nodes or by increasing $\Delta_{\rm min}$), it will lead to more
zero-energy quasiparticles in a gap with $d$-wave symmetry, for
example \cite{Borkowski}.  Both Nohara {\it et al.}  \cite{Nohara} 
and Yokoya {\it et al.}  \cite{Yokoya} have
interpreted their data on pure and impure YNi$_2$B$_2$C 
(specific heat and photoemission spectroscopy, respectively)
in terms of an anisotropic $s$-wave gap.


On the question of a phonon mechanism, various authors
\cite{Eliashberg} have argued that a standard Eliashberg analysis of
$H_{c2}(T)$ and other data leads to a quantitatively satisfactory and
consistent description of \lu\ in terms of the measured phonon
spectrum and a largely isotropic gap function.  It remains to be seen
whether such an analysis survives the inclusion of a very
anisotropic gap.

In conclusion, we have shown the presence of highly delocalized
quasiparticles throughout the vortex state of \lu.  The quasiparticle
transport grows as a function of magnetic field in the same way as it
does in UPt$_3$, an unconventional superconductor known to have a line
of nodes in the gap, and not at all like $s$-wave superconductors.  We
conclude that the gap function of \lu\ must have nodes in the gap, or
at least deep minima.  More work is necessary to determine precisely
the location of these nodes/minima.  Such pronounced gap anisotropy is
unprecedented in phonon-mediated superconductivity, raising the
question of whether phonons are indeed responsible for Cooper pairing
in borocarbide superconductors, as has traditionally been thought.






We are grateful to Z. Te\u{s}anovi\'{c}, S. Vishveshwara, T. Senthil,
M.P.A. Fisher, M. Walker and V. Kogan for stimulating discussions.
This work was supported by the Canadian Institute for Advanced
Research and funded by NSERC of Canada. 
Ames laboratory is operated for the U.S.
Department of Energy by Iowa State University under contract No.
W-7405-ENG-82.


\end{document}